\documentclass[a4paper, amsfonts, amssymb, amsmath, reprint, showkeys, nofootinbib, twoside]{revtex4-1}
\usepackage[english]{babel}
\usepackage[utf8]{inputenc}
\usepackage[colorinlistoftodos, color=green!40, prependcaption]{todonotes}
\usepackage{amsthm}
\usepackage{mathtools}
\usepackage{physics}
\usepackage{xcolor}
\usepackage{graphicx}
\usepackage[left=23mm,right=13mm,top=35mm,columnsep=15pt]{geometry} 
\usepackage{adjustbox}
\usepackage{placeins}
\usepackage[T1]{fontenc}
\usepackage{lipsum}
\usepackage{csquotes}
\usepackage[pdftex, pdftitle={Article}, pdfauthor={Author}]{hyperref} 
\begin{document}
\title{Titanium and titanium oxides at the K- and L-edges: validating theoretical calculations of X-ray absorption and X-ray emission spectra with measurements}

\author{Karina Bzheumikhova}
    \email[Correspondence email address: ]{karina.bzheumikhova@ptb.de}
    \affiliation{Physikalisch-Technische Bundesanstalt, Abbestra{\ss}e 2-12, 10587 Berlin, Germany}

\author{John Vinson}
    \affiliation{Material Measurement Laboratory, National Institute of Standards and Technology, Gaithersburg, MD 20899}
\author{Rainer Unterumsberger}
    \affiliation{Physikalisch-Technische Bundesanstalt, Abbestra{\ss}e 2-12, 10587 Berlin, Germany}
\author{Malte Wansleben}
    \affiliation{Physikalisch-Technische Bundesanstalt, Abbestra{\ss}e 2-12, 10587 Berlin, Germany}
\author{Claudia Zech}
    \affiliation{Physikalisch-Technische Bundesanstalt, Abbestra{\ss}e 2-12, 10587 Berlin, Germany}
\author{Kai Sch{\"u}ler}
    \affiliation{Physikalisch-Technische Bundesanstalt, Abbestra{\ss}e 2-12, 10587 Berlin, Germany}
\author{Yves Kayser}
    \affiliation{Physikalisch-Technische Bundesanstalt, Abbestra{\ss}e 2-12, 10587 Berlin, Germany}
\author{Philipp H{\"o}nicke}
    \affiliation{Physikalisch-Technische Bundesanstalt, Abbestra{\ss}e 2-12, 10587 Berlin, Germany}
\author{Burkhard Beckhoff}
    \affiliation{Physikalisch-Technische Bundesanstalt, Abbestra{\ss}e 2-12, 10587 Berlin, Germany}

\date{\today} 

\begin{abstract}
Using well-calibrated experimental data we validate theoretical X-ray absorption spectroscopy (XAS) as well as X-ray emission spectroscopy (XES) calculations for titanium (Ti), titanium oxide (TiO), and titanium dioxide (TiO$_2$) at the Ti K- and L-edges as well as O K-edge. XAS and XES in combination with a multi-edge approach offer a detailed insight into the electronic structure of materials since both the occupied and unoccupied states, are probed. The experimental results are compared with \textit{ab initio} calculations from the {\sc ocean} package which uses the Bethe-Salpeter equation (BSE) approach. Using the same set of input parameters for each compound for calculations at different edges, the transferability of the {\sc ocean} calculations across different spectroscopy methods and energy ranges is validated. Thus, the broad applicability for analysing and interpreting the electronic structure of materials with the {\sc ocean} package is shown.
\end{abstract}

\maketitle

\section{Introduction}
In the present study we use titanium and several of its oxides as model systems for early 3d transition metals, and employ X-ray absorption spectroscopy (XAS) and X-ray emission spectroscopy (XES) at different energy ranges (the Ti and O K edges and the Ti $L_{II,III}$ edges) to validate {\it ab initio} calculations using the {\sc ocean} code \cite{vinson_2011, vinson_2022, gilmore_2015}. The study of early 3d transition metal compounds by means of XAS and XES provides information on the electronic structure which has an influence on material properties like charge transport and optical properties relevant in many fields including nanotechnologies and energy storage materials \cite{baermann_2021, sibum_2017, peng_2017, krishnamoorthy_2017}.

The electronic structure of these titanium compounds has been previously well studied \cite{kung_1989, farges_1997, cabaret_2010, mastelaro_2018, groot_2021}. That is why titanium and its oxides are excellent candidates to test and validate theoretical calculation methodologies. Therefore, we use this knowledge combined with physically traceable measurement techniques with complementary discrimination and sensitivity capabilities to provide a reliable validation of {\it ab initio} calculations based on the Bethe-Salpeter equation (BSE) approach. We show that moderately complex structures such as early transition metal like titanium can be well de\-scribed using the BSE approach. The same set of {\sc ocean} input parameters for each compound is used for the calculations, which are then compared to XAS and XES data from both the K- and L-edges. The predictive power of the {\sc ocean} code, which explicitly treats the interactions between the photo-electron and core-hole using the BSE approach, has been shown for specific cases in the past \cite{vinson_2014, andrle_2020, singh_2018}. 
Wansleben, et al., provides another example of a successful comparison between {\sc ocean} calculations with experimental XES and XAS data for iron sulfur compounds \cite{wansleben_2020}.
This present work continues these investi\-gations by validating the code for a multi-edge analysis of an early transition metal.

X-ray spectroscopy techniques at different edges and energies can be used to probe distinct electron transitions and thus different parts of the electronic structure. On the other hand, the discrimination capability and sensitivity of XAS and XES at different edges vary significantly \cite{groot_2001, groot_2005, reinhardt_2009}. Quite often available instrumentation or experimental boundary conditions regarding the sample and its environment influence the type of measurement which can be conducted. In this context a prior consideration based on calculations which need to be validated, using studies as presented in this work, will be useful to evaluate the discrimination capability which can be expected during the experiment. It is shown that this type of consideration can be realized for different energy ranges and spectroscopic techniques. XAS probes the unoccupied part of the electronic structure of the system. The oxidation state of an element can be determined with XAS at the K edge by analysing the edge position and the pre-edge, which can be difficult to measure and interpret. Indeed, the pre-edge of a transition metal displays the dipole-forbidden but quadrupole-allowed 1s to 3d transitions \cite{farges_1997, cutsail_2022}. While dipole-forbidden transitions can be observed in weak quardupole and mixed transitions in the pre-edge, the involved states can be observed in dipole-allowed transitions at the L edge. XES, on the other hand, probes the occupied part of the electronic structure. Specifically, the K$\beta$ spectroscopy is of interest since the K$\beta$ satellites K$\beta_{2,5}$ and K$\beta^{''}$ are sensitive to ligand bonds of transition metals \cite{bauer_2014, kowalska_2015, schwalenstocker_2016}. These are, however, challenging to measure due to low intensity of around $2\%$ relative to the K$\beta_{1,3}$ \cite{mandic_2009}, whose high-energy tail is overlapping with the satellites. We successfully complete this analysis of the ligand bonds by including oxygen K-edge experiments, which probe the core-level transitions of the oxygen atom.

A careful comparison of theoretical and experimental data requires an understanding of different factors that might have influenced the data. In order to move towards reliable and quantitative investigations, we use calibrated and well characterized instrumentation which allows us to differentiate between instrumental, experimental and physical contributions. For comparing experimental spectra to calculations of the electronic structure it is important that the spectra are free from the instrumental and experimental contributions. These influences can partially be mitigated. Instrumental resolutions, efficiencies, and uncertainties can be optimized \cite{beckhoff_2008, beckhoff_2009, mueller_2009_2} and energy scales can be calibrated, optimally in a traceable manner such that a transfer to and intercomparability of different experiments with regard to the instrumentation used can be realized. To start with, this is done for our measurements by using instrumentation that has been calibrated in terms of its energy scale in a physically traceable manner and response function at selected incident and emission photon energies. The calibration of the energy scales of monochromatized synchrotron radiation allows to use a common energy scale for XAS and XES measurements involving the creation of an electron vacancy in the same shell, which are necessarily collected with different instrumentation under the premise that the spectrometer used for the XES experiments can be calibrated using elastically scattered radiation. This aspect is crucial for the comparability and validation of the different calculations for XAS and XES. 

\section{Experimental setup and measurement procedure}
All experiments were conducted in the Physikalisch-Technische Bundesanstalt (PTB) laboratory at the electron storage ring BESSY II \cite{klein_2002, beckhoff_2000}. XAS measurements around the K-edge were realized at the four-crystal-monochro\-mator (FCM) beamline \cite{krumrey_1998, krumrey_2001}, while measurements around the L-edges were realized at the plane-grating monochromator (PGM) beamline \cite{senf_1998} for undulator radiation. The beamlines provide tunable radition in the X-ray photon energy ranges between 1.75~keV and 10~keV and 78~eV and 1860~eV, respectively, with high spectral purity and photon flux. The energy scale of the FCM beamline has been calibrated using back-reflection of single-crystals \cite{krumrey_1998, krumrey_2001}. The energy resolution is $\approx 0.38$~eV around the titanium K edge and the uncertainty is around 1.0~eV. The PGM beamline is calibrated from the absorption of well-known vibrational resonances of noble gases. The resolution of the PGM beamline around the titanium L edge is $\approx0.23$~eV and an uncertainty is around 0.5~eV \cite{senf_1998}.

The measurements shown in this work are complemented by XES K-edge measurements and calculations first pub\-lished by Wansleben {\it et al.} \cite{wansleben_2019} as well as XES and XAS mea\-surements at the Ti L-edge published by Unterums\-berger {\it et al.} \cite{unterumsberger_2018}. 
Together they form a complete picture of the near-edge x-ray emission and absorption which we use to assess the validity of the {\sc ocean} code across various core levels. 

X-ray absorption measurements were conducted by both detecting the induced fluorescence radiation with a silicon drift detector (SDD) \cite{krumrey_2006, scholze_2009} as well as transmission measurements \cite{unterumsberger_2018}. A photodiode is used to determine the incident photon energy dependent variations in the photon flux. The samples were inserted in an ultra-high vacuum environment at an incident and takeoff angle of $\theta = 45^{\circ}$ \cite{beckhoff_2008}. The instrumental influence by means of the response functions is quantified such that the different contributions in the measured spectra can be accurately discriminated. The diodes used for transmission measurements as well as the response function of the SDD used for the fluorescence radiation measurements are calibrated. 

XES experiments around the Ti K edge were conducted at the dipole white light (DWL) radiation beamline \cite{thornagel_2001} with a polychromatic excitation spectrum originating from a 1.3~T bending magnet. To reduce background radiation caused by scattering the polychromatic excitation spectrum was attenuated by a 2-$\mu$m-thick aluminum filter \cite{wansleben_2019}. The measurements were conducted using a von Hamos spectrometer \cite{anklamm_2014} based on two full-cylinder Bragg crystals, each consisting of 40-$\mu$m-thick highly annealed pyrolytic graphite (HAPG) mosaic crystals. The use of two crystals instead of one increases the resolving power of the spectrometer \cite{hohlwein_1988}. Further details on the instrumentation and measurement details of the titanium XES measurements around the K edge can be found in the work of Wansleben {\it et al.} \cite{wansleben_2019}.

The XES spectra around the L edges as well as oxygen K edge were collected at the PGM beamline using a wavelength-dispersive spectrometer (WDS) \cite{nordgren_1989, mueller_2009_2, unterumsberger_2015} based on the Rowland circle geometry. The optical source, defined by means of an entrance slit, the reflective grating, and the detecting charge coupled device (CCD) detector are all positioned on the Rowland circle which is defined by the grating curvature. A vertical slit placed between the entrance slit and the grating collimates horizontally so that the detector is not illuminated outside of its active area in the non-dispersive direction. Additionally, a horizontal slit is restricting the illuminated area of the grating in the dispersive direction defining the solid angle of acceptance of the WDS. 

For the validation of the broad applicability of the {\sc ocean} code a set of samples was used that is well known and has been studied with X-ray spectrometry techniques \cite{unterumsberger_2018}. Three different samples were used: Ti, TiO, and TiO$_2$. The used samples are titanium oxides on a thin silicon nitride window which is placed on a silicon wafer. Different titanium oxidation states were achieved by varying the amount of oxygen using ion beam sputter deposition (IBSD), measured in standard cubic centimeters per minute. The oxidation state of the samples depends on the oxygen flow rate. 

\section{Theoretical Modeling} \label{theor}
The XES and XAS calculations have been carried out throughout this work using the {\sc ocean} code \cite{vinson_2011, gilmore_2015}. The first-principle code {\sc ocean} calculates core edge spectroscopy including XAS, XES, RIXS, and non-resonant X-ray scattering (NRIXS). {\sc ocean} input includes parameters on the atomic structure, photon excitation information, pseudopotentials generated using the ONCVPSP code \cite{hamann_2013} for the density functional theory (DFT) calculation, and specific convergence thresholds for the calculations. For each compound, no input parameters are changed when switching between edges or absorption and emission. Thus, the comparison of calculated results with experimental results from different energy ranges (different edges) and different types of measurements (XAS, XES) allows for a broad validation of the code. 

The general issue with 3d transition metals is the partially filled 3d-sub-shell. Early transition metals have mostly unoccupied 3d-bands, which is well described within DFT, whereas for later transition metals local or semi-local density functionals fail to sufficiently localize 3d- or 4f-electrons and underestimate the strength of the Coulomb repulsion between electrons of opposite spin in the same orbital.

Norm-conserving pseudopotentials and the local-density approximation to the exchange-correlation functional were chosen in this work. The atomic structures were taken from the Crystallography Open Database (Ti \cite{glav_1992}, TiO \cite{yu_1974}, TiO$_2$ \cite{wyckoff_1965}). The plane-wave cutoff is used for the DFT calculation to truncate the basis and has been chosen as stated in Table \ref{table:parameters}. The \textbf{k}-mesh grid of the crystal momentum for the BSE was chosen to be sufficiently high within a reasonable computation time. The number of conduction and screening bands is chosen to be sufficiently high so that the number of wave functions included up to some energy above the Fermi level are well represented. Sufficiently high numbers of conduction and screening bands were achieved by increasing the respective parameter until the difference in the energy scale of the system was less than 0.01~eV. The final set of parameters is shown in Table \ref{table:parameters}. 

\begin{table}
    \small
    \begin{tabular*}{0.48\textwidth}{@{\extracolsep{\fill}}llll}
        \hline
         & Ti & TiO & TiO$_2$  \\
        \hline
        plane-wave cutoff / Ry. & 200 & 150 & 150 \\
        k-mesh & 12$\times$12$\times$10 & 8$\times$8$\times$8 & 12$\times$12$\times$8 \\ 
        number of conduction bands  & 400 & 400 & 500 \\ 
        number of screening bands & 700 & 397 & 1235 \\ \hline
    \end{tabular*}
    \caption{{\sc ocean} calculation convergence parameters. The plane-wave cutoff (given in Rydberg) and the conduction bands number are used for the DFT calculation, whereas the k-mesh and screening band numbers are used for the BSE calculation.}
    \label{table:parameters}
\end{table}

\section{Results and Discussion}
In the following we present the accumulated results for the XAS and XES measurements around the K and L edges for the three titanium compounds Ti, TiO, and TiO$_2$. It is important to note, that the presented calculations are done in terms of band structure calculations on crystal structures of solids. However, for simplicity and a convenient comparison with theoretical data, the molecular-orbital (MO) model has been chosen to describe the features in the spectra \cite{fischer_1970, tsutsunami_1977}. In the MO representation titanium and oxygen build molecular orbitals which include valence electrons of titanium $4s$ and $3d$, and oxygen $2p$ as well as stronger bound $2s$ oxygen electrons and the unoccupied titanium $4d$ state. TiO and TiO$_2$ differ  mostly in the resulting $2t_{2g}$ state. While TiO has two electrons in this state, TiO$_2$ has an unoccupied $2t_{2g}$ state.

When comparing experimental and calculated spectra the focus lies on peak positions and shapes. The deviations in peak shape between the two are the result of both experimental and theoretical factors. One of the experimental aspects is the resolution of the instrumentation. In all the presented experimental data the resolution of the beamline and the chosen spectrometers is known. Thus, it was applied to the calculated data as an energy independent broadening, which is sufficient in the given relatively small energy ranges for each individual experiment. The summarized broadening resulting from the instrumentation can be found in Table \ref{table:resolution} \cite{krumrey_2001, mueller_2009_2, wansleben_2019, senf_1998}. Additionally, a theoretical broadening due to the core-hole lifetime was applied to the calculated spectra. For the titanium K edge this is 0.94 eV, for the L edge 0.24 eV (L$_{II}$ edge value), and for the oxygen K edge 0.13 eV \cite{zschornack_2006}. It is important to note that additional broadening effects like super-Coster-Kronig decay processes and the radiative Auger effect are neglected in this first order treatment \cite{enkisch_2004, taguchi_2013}.

The experimental data have an uncertainty from the respective beamline and spectrometer energy scales' uncertainties. For the measured spectra this arises due to the energy scale of the used beamlines and spectrometers. The FCM beamline energy is calibrated using a back-reflection of single crystals \cite{krumrey_2001} while the PGM beamline is calibrated using resonances of noble gases \cite{senf_1998}. This way a transferability of the energy scale is provided between different measurements. In addition, the emission energy scale for the XES measurements is defined using the elastic scattering of a material and thus transferred from the energy scale of the beamline. The energy scale of the XES K measurements is calibrated using previous measurements at the FCM beamline. There, using the monochromatic energy available the energy scale has been calibrated using the elastic scattering and then transferred to the DWL beamline. Based on continuous calibrations of the beamlines throughout the years, we determined an uncertainty of the two beamlines. For the PGM beamline the uncertainty of the energy scale is around 0.5 eV, whereas for the FCM beamline it is around 1 eV.

{\sc ocean} does not calculate an absolute energy scale. Therefore, one of the calculations is used to align the energy scale with the experimental energy scale by determining either a significant feature position or the edge position and using this offset for all other calculated data. This is done once for the Ti K edge, the Ti L edge, and the O K edge. Meaning that the energy alignment is done only once for each edge and used throughout the respective XAS and XES spectra for all three compounds. In this way the relative positions calculated with {\sc ocean} remain intact and can be validated against the calibrated experimental data. 

\begin{figure}[ht]
    \centering
    \includegraphics[width=.8\linewidth]{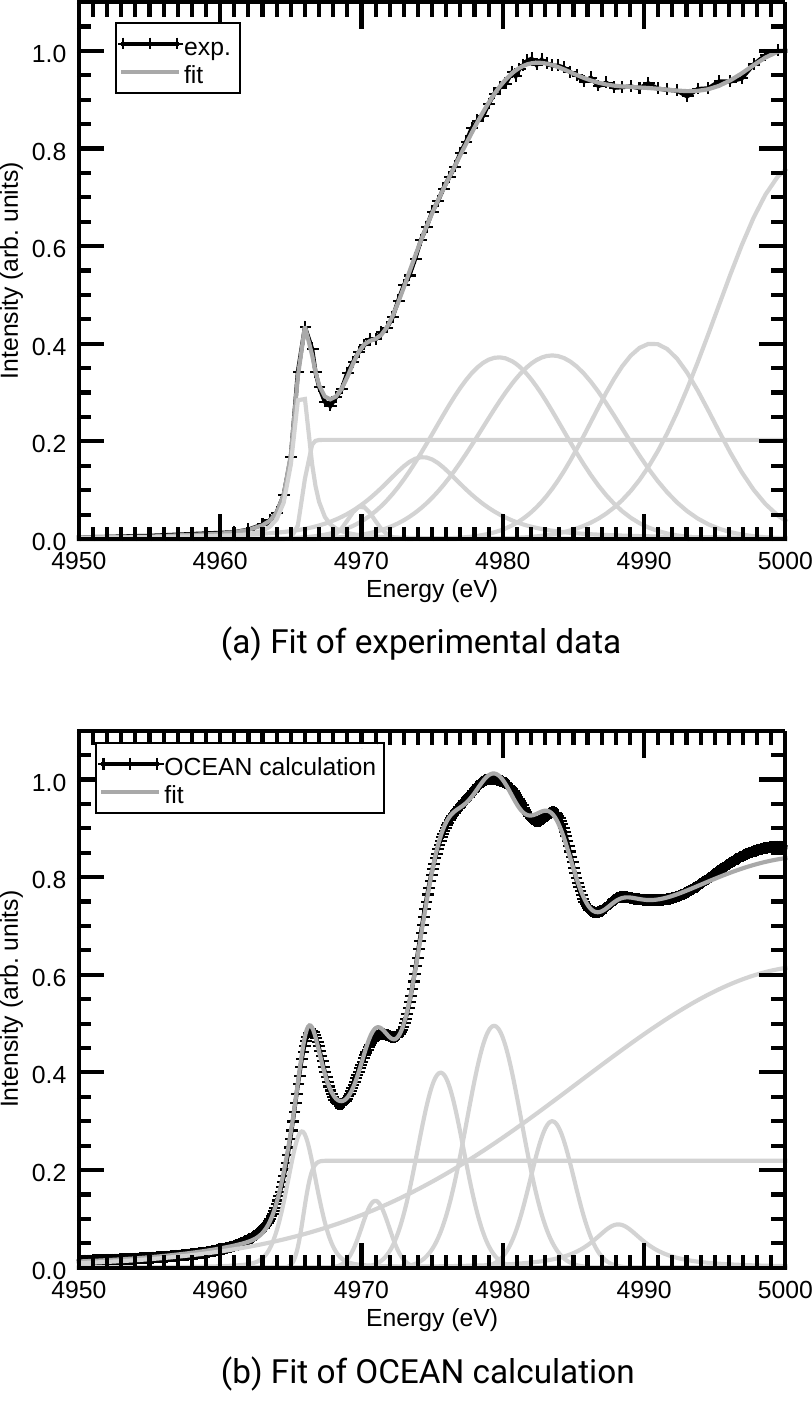}\quad
    \caption{Fit of K-edge XAS results of Ti as measured (a) and calculated with {\sc ocean} (b). For fitting of the edge an error function was used whereas for the peaks a Voigt profile consisting of a Gaussian convolved with Lorentzian was used. }
    \label{fig:xas_k_fit}
\end{figure}

For further analysis of the comparison of spectra regarding the peak positions between experimental data and theoretical calculations as well as the extraction of the pre-edge peaks, a fit of all spectra was performed using Voigt functions and additional step functions in the case of XAS \cite{zech_2021}. A Voigt function consists of a Gaussian and a Lorentzian, which were chosen to represent physical conditions. The Lorentzian broadening was able to change around the corresponding natural line widths \cite{zschornack_2006}. The Gaussian was allowed to be within a reasonable experimental boundary defined by the minimum known instrumental resolution as can be found in Table \ref{table:resolution}. An example of a fit is shown in Fig. \ref{fig:xas_k_fit}. The summary of all the peak position deviations between calculation and experimental data can be found in Table \ref{table:summary}, which is explained in detail in the following sections regarding the different measurements.

Deviations in the peak number are mostly due to broadening in the experiment, which precludes resolving all the features present in the calculated spectra. However, since the experimental data does not resolve all peaks, for the comparison in Table \ref{table:summary} the respective peaks from the {\sc ocean} fit were combined and the maxima were used for the comparison. 

\begin{table}[ht]
    \small
    \begin{tabular*}{0.48\textwidth}{@{\extracolsep{\fill}}llll}
        \hline
        Instrum. & Res. power E/$\Delta E$ & Resol. (eV) & Edge \\
        \hline
        FCM & 11800 & 0.38 & Ti K \\ 
        PGM & 2000 & 0.23 & Ti L $\&$ O K \\
        von & 2700 & 1.8 & Ti K \\
        WDS  & 1100 & 0.4 & Ti L $\&$ O K\\ 
        \hline
    \end{tabular*}
    \caption{Summarized resolving power and resolutions from the beamlines and the spectrometers used \cite{krumrey_2001, mueller_2009_2, wansleben_2019}. The overall broadening resulting for each measurement was applied to the calculated data as a Gaussian broadening.}
    \label{table:resolution}
\end{table}

\begin{table}
    \centering
    \includegraphics[width=1.0\linewidth]{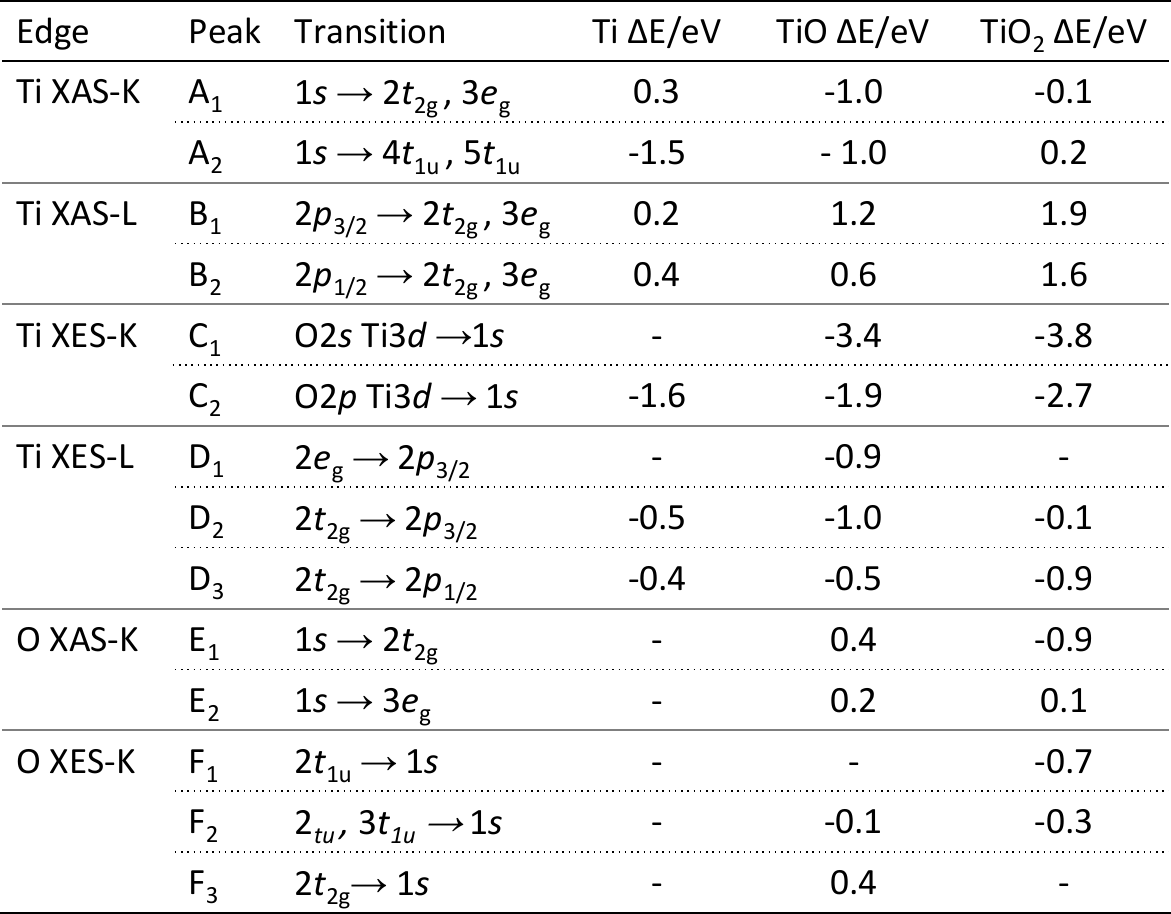}\quad
    \caption{Summarized differences between calculated and measured peak positions.}
    \label{table:summary}
\end{table}

\subsection{Titanium K-edge XAS}
\begin{figure}[ht]
    \centering
    \includegraphics[width=.8\linewidth]{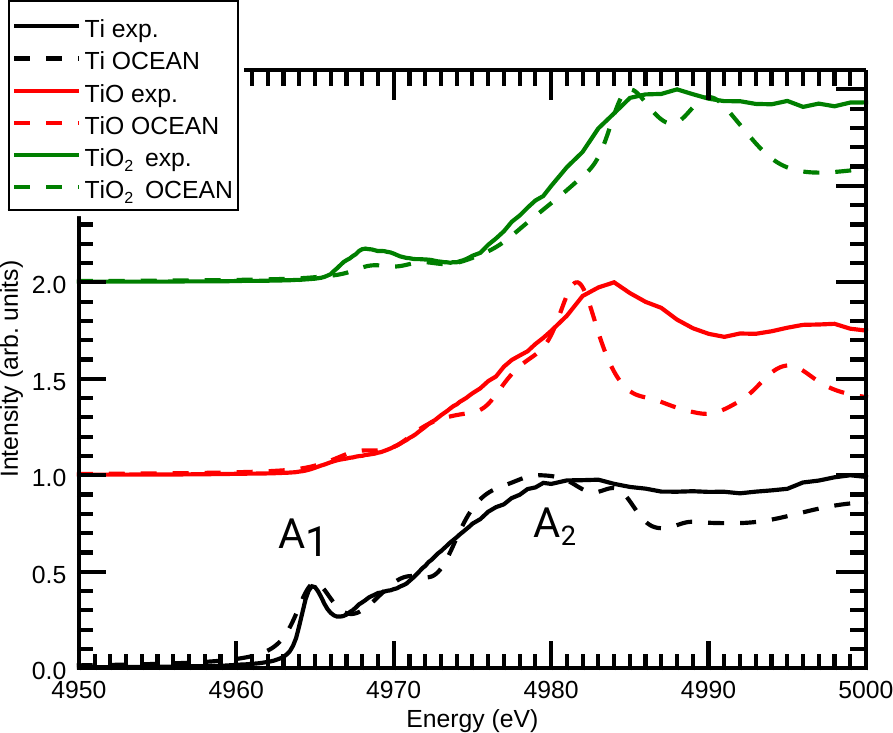}\quad
    \caption{Comparison between measured (solid lines) and calculated (dashed lines) spectra of K-edge XAS for Ti, TiO, and TiO$_2$. The spectra are normalized to the respective maximum intensity and offset vertically for clarity.}
    \label{fig:xas_k}
\end{figure}

\begin{figure}[ht]
    \centering
    \includegraphics[width=.8\linewidth]{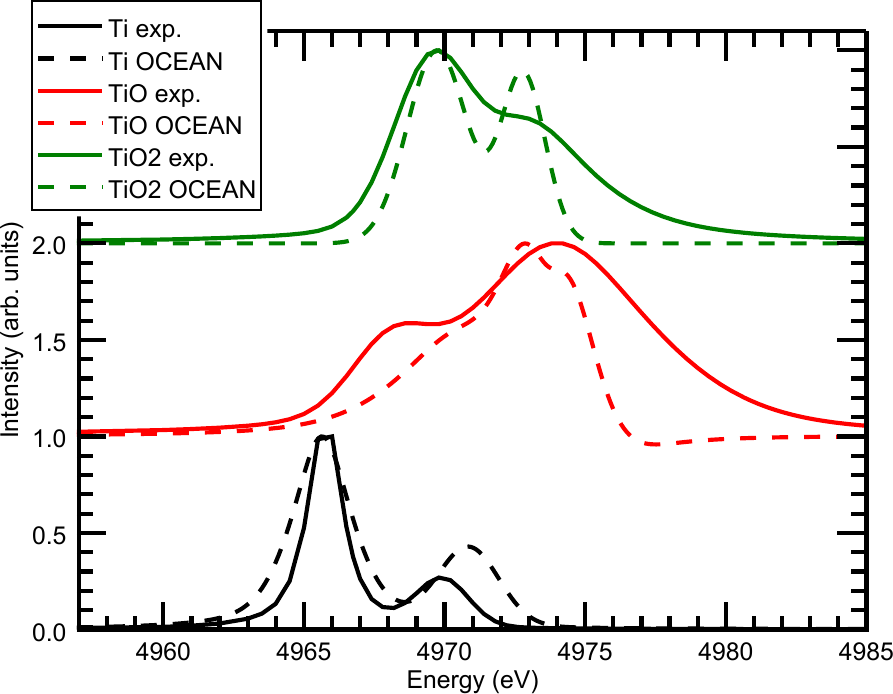}\quad
    \caption{Comparison between measured (solid lines) and calculated (dashed lines) spectra of the extracted pre-edge. The spectra are normalized to the respective maximum intensity and offset vertically for clarity.}
    \label{fig:xas_k_pre}
\end{figure}

The obtained XAS results at the K-edge are shown in Fig. \ref{fig:xas_k} as the normalized intensity as a function of the monochromatic incident photon energy. The spectra are presented normalized to the maximum intensity and offset vertically for clarity. XAS probes the transition into the unoccupied states, which results in the characteristic K-edge features A$_1$ and A$_2$. The pre-edge feature A$_1$ can be assigned to the $1s \rightarrow 2t_{2g}$ and $1s \rightarrow 3e_g$ quadrupole transitions \cite{sorantin_1992} while the feature A$_2$ are the $1s \rightarrow 4t_{1u}$ and $1s \rightarrow 5t_{1u}$ transitions. Differences in the three compounds can be analyzed through the shape and energy shift differences between these features. With higher oxidation state the overall spectra shift to higher energies by about 4~eV and 7~eV respectively. This is accurately represented in the {\sc ocean} calculation. Note that the same absolute energy shift was applied to all spectra to match the experimental data.

The feature A$_2$ is broad in the experimental data and can not be resolved as several features. However, a slight change in the shape of the feature can be seen between the three compounds. TiO seems to have the most pronounced shape while Ti has the broadest peak A$_2$. Despite the efforts of including broadening into the calculation, the peaks in the experimental data remain broader than the calculated results. An additional factor which was not taken into account in this work is the vibrational disorder. All calculations have been carried out using a unit cell, which ignores the fact that disorder might introduce an additional broadening to the whole system \cite{vinson_2016}. Additional calculations using a super cell can be carried out, but have not been included in this work.

Fig. \ref{fig:xas_k_pre} shows a comparison of the extracted pre-edge for all three materials in comparison between experimental data and the respective {\sc ocean} calculation. The extraction was done using the previously mentioned fitting of the spectra and separating the pre-edge from the edge step and the excitonic peaks at the edge. The pre-edge region of a metal compound can be used to extract information on the coordination number, the oxidation state as well as the spin-state of the absorbing atom \cite{yamamoto_2008, cabaret_2010}. Cabaret {\it et al.} \cite{cabaret_2010} showed that the pre-edge region of TiO$_2$ can be assigned to both quadrupolar $t_{2g}$ as well as mixed dipolar and quadrupolar $e_g$ transitions. The experimental and theoretical data have a good agreement concerning the pre-edge peak intensities, ratios and positions. 

\subsection{Titanium L-edge XAS}
\begin{figure}[ht]
    \centering
    \includegraphics[width=.8\linewidth]{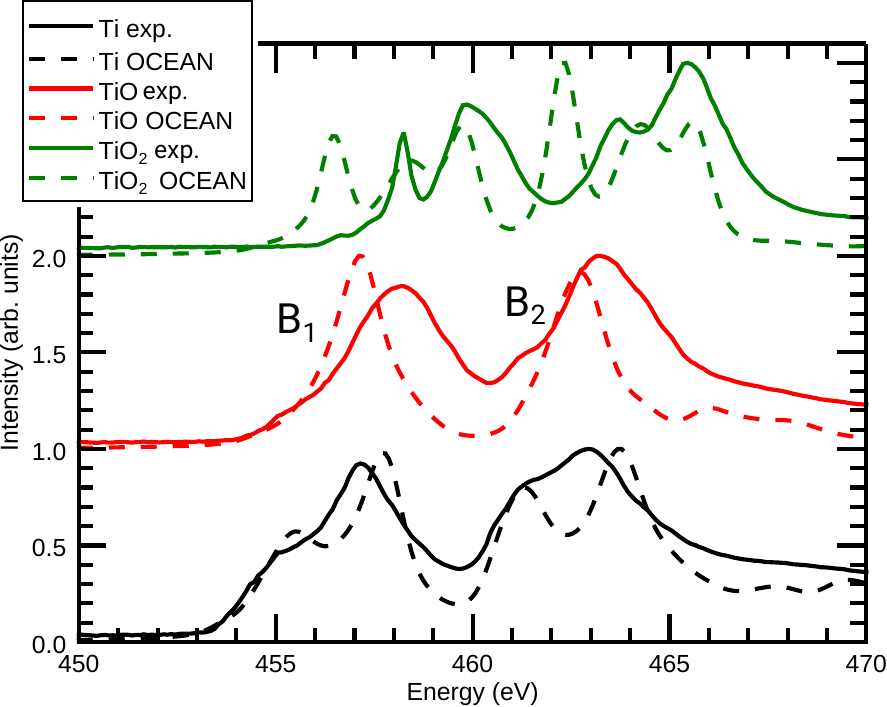}\quad
    \caption{Comparison between measured (solid lines) and calculated (dashed lines) spectra of L-edge XAS for Ti, TiO, and TiO$_2$. The spectra are normalized to the respective maximum intensity and offset vertically for clarity.}
    \label{fig:xas_l}
\end{figure}

The obtained XAS results at the L edges are shown in Fig. \ref{fig:xas_l} as the normalized intensity and as a function of the monochromatic incident photon energy. The previously mentioned 2t$_{2g}$ molecular orbital is where the three compounds differ the most. While at the K edge, this state is only involved in the forbidden quadrupole transition, at the L edge it is a dipole transition. Thus, the involved transitions have relatively high intensity among the observable transitions. This aspect makes L-edge spectroscopy generally more sensitive for the characterization of titanium oxides. The L edges consist of two main features: B$_1$ representing the L$_{III}$ and B$_2$ representing the L$_{II}$ edge transitions \cite{unterumsberger_2015}. Similar to the K-edge XAS, the spectra shift with higher oxidation state towards higher energies. However, the calculated data has a discrepancy within TiO and TiO$_2$ calculation regarding the overall energy shift. While the general form of the spectra has a very good agreement in terms of peak form and energy differences between peaks, {\sc ocean} appears to predict the energy offset for the ligand bond wrongly. However, the discrepancy capability regarding different chemical species at the L-edges is not manifested in the energy shift, but rather in the spectra form and the branching ratio between the L$_{III}$ and L$_{II}$ edges. {\sc ocean} captures the branching ratio between the L$_{III}$ and L$_{II}$ edges correctly. Based on the 2$p$ occupation and the available 3$d$ states the statistical value of the branching ratio would be 2:1. However, this is not the observation in measured data. Shirley {\it et al.} \cite{shirley_2005} as well as Laskowski {\it et al.} \cite{laskowski_2010} showed that this is due to mixing between the excitations from 2$p_{1/2}$ and 2$p_{3/2}$ states which can be observed by setting the exchange term of the BSE Hamiltonian to zero. Our {\sc ocean} calculations support this theory and are in agreement with the measured data. Both, calculations and experimental data, have a branching ratio around 1:1. 
 
\subsection{Titanium K-edge XES}
\begin{figure}[ht]
    \centering
    \includegraphics[width=.8\linewidth]{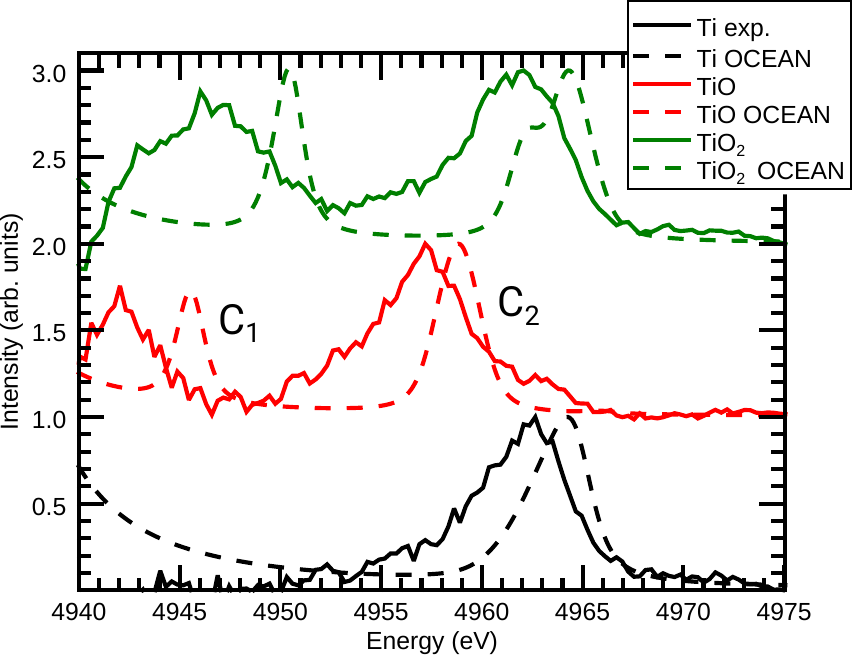}\quad
    \caption{Comparison between measured spectra (solid lines) from Wansleben {\it et al.} \cite{wansleben_2019} and calculated (dashed lines) spectra of XES for Ti, TiO, and TiO$_2$ above the titanium K-edge. The spectra are normalized to the respective maximum intensity and offset vertically for clarity.} 
    \label{fig:kbeta}
\end{figure}

\begin{table}
    \centering
    \includegraphics[width=1.0\linewidth]{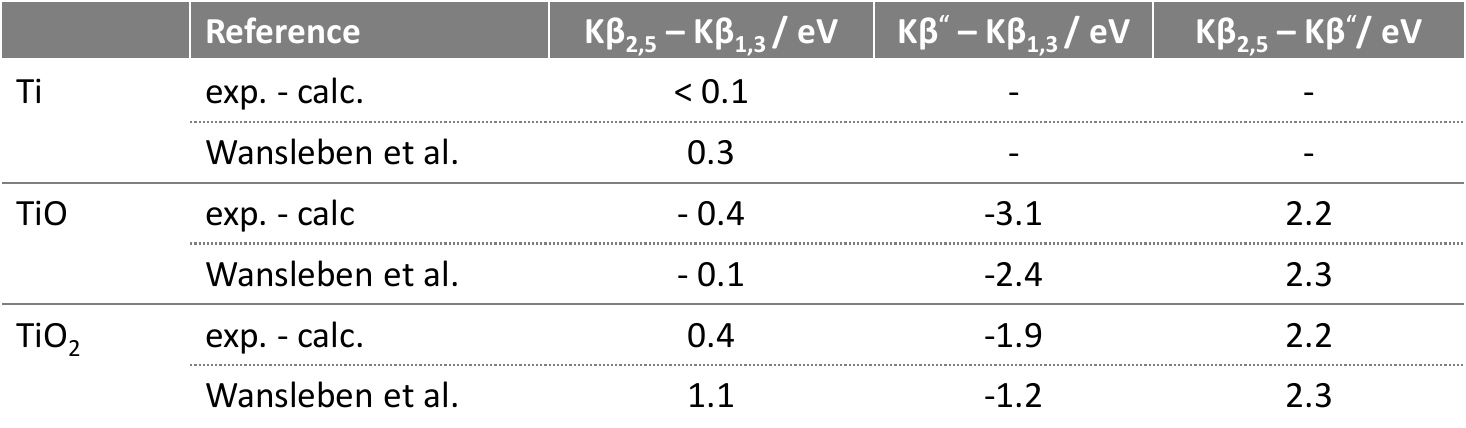}\quad
    \caption{Summarized differences between XES K-edge calculated and measured K$\beta^"$ and K$\beta_{2,5}$ peaks relative to the K$\beta_{1,3}$ line and each other in comparison to data taken from Wansleben {\it et al.}. \cite{wansleben_2019}.}
    \label{table:summary_xesk}
\end{table}

An extensive study of the K-edge, in particular K$\beta$ spectroscopy, including {\sc ocean} calculations has been done by Wansleben et al. \cite{wansleben_2019} (Fig. \ref{fig:kbeta}). Therefore, we limit our discussion of the K-edge XES to the differences between that work and this study. Although the same structure and the same set of parameters was chosen, the calculations were carried out with a different set of pseudopotentials. While Wansleben {\it et al.} chose pseudopotentials of the Fritz Haber Institute (FHI) from the {\sc Quantum ESPRESSO} website, this work uses pseudopotentials created with the {\sc ONCVPSP} code, as stated in section \ref{theor}. Our results and the results from Wansleben {\it et al.} are presented in Table \ref{table:summary_xesk}. The results are fairly similar with a few exceptions for the K$\beta^{''}$-K$\beta_{1,3}$ differences between the peaks. It is important to note, that the reason for this, next to differences in the calculation, might be due to differences in the fitting and extraction of the spectra.

\subsection{Titanium L-edge XES}
\begin{figure}[ht]
    \centering
    \includegraphics[width=.8\linewidth]{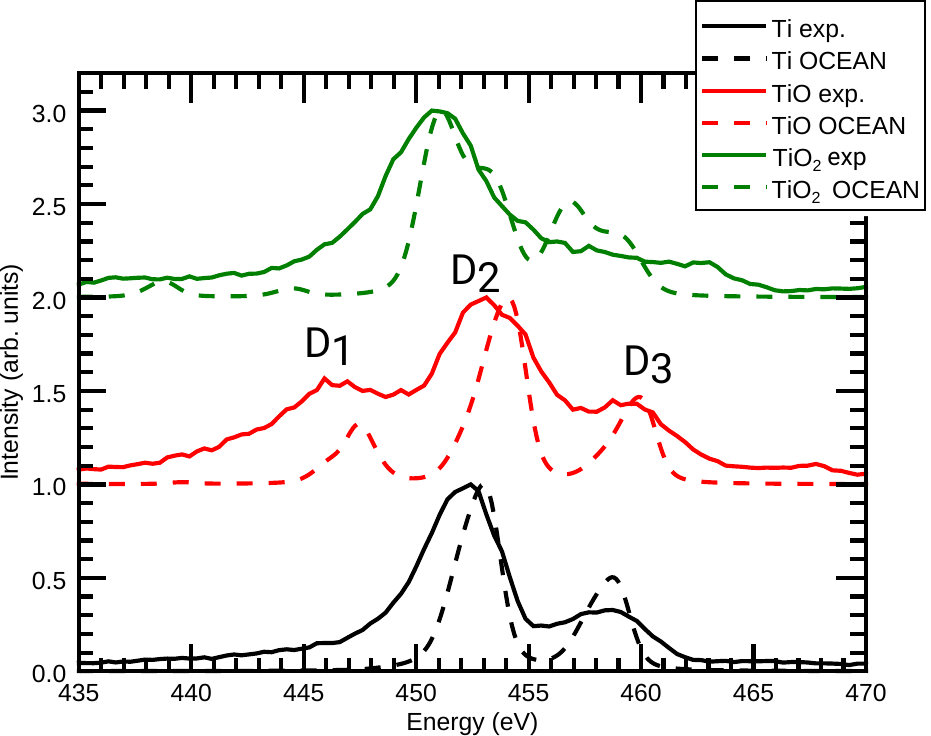}\quad
    \caption{Comparison between measured (solid lines) and calculated (dashed lines) spectra of XES for Ti, TiO, and TiO$_2$ above the titanium L-edge. The spectra are normalized to the respective maximum intensity and offset vertically for clarity.}
    \label{fig:xes_l}
\end{figure}

The results of the comparison of calculated data to experimental data of XES around the L-edges are presented in Fig. \ref{fig:xes_l}. The metallic Ti compound shows two distinct features related to the L$\alpha$ and L$\beta$ transitions \cite{unterumsberger_2018}. With higher oxidation states transitions from molecular orbitals are important to consider. The feature D$_1$ is related to the $2e_g \rightarrow 2p_{3/2}$, the feature D$_2$ to the $2t_{2g} \rightarrow 2p_{3/2}$, and D$_3$ to the $2t_{2g} \rightarrow 2p_{1/2}$ transition. The {\sc ocean} spectra show a very good agreement of all three compounds regarding peak position, peak number, and energy shifts between the peaks. The occurrence of the peak D$_1$ for the oxides is correctly predicted as well as the peak ratios between the different compounds, which is relevant for the chemical speciation capability. 

\subsection{Oxygen K-edge}
\begin{figure}[ht]
    \centering
    \includegraphics[width=.8\linewidth]{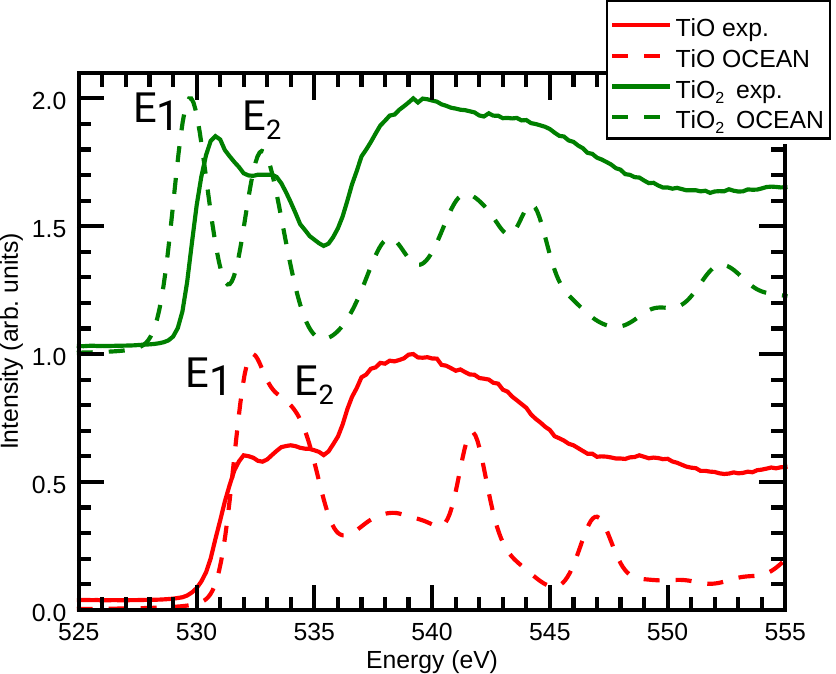}\quad
    \caption{Comparison between measured (solid lines) and calculated (dashed lines) spectra of O K-edge XAS for TiO, and TiO$_2$. The spectra are normalized to the respective maximum intensity and offset vertically for clarity.}
    \label{fig:o_k_nex}
\end{figure}
In addition to titanium K- and L-edge data, the O K edge was analyzed for the titanium oxide compounds. The XAS K-edge results are shown in Fig. \ref{fig:o_k_nex}. These display a significant change around the $\pi^*$-resonance with a shift in energy of 1.2~eV as well as a change regarding the form. Although the TiO$_2$ has a slightly larger offset in the energy between the calculation and experimental data, both, TiO and TiO${_2}$, reveal a clear discrimination opportunity around the oxygen K edge. 

\begin{figure}[ht]
    \centering
    \includegraphics[width=.8\linewidth]{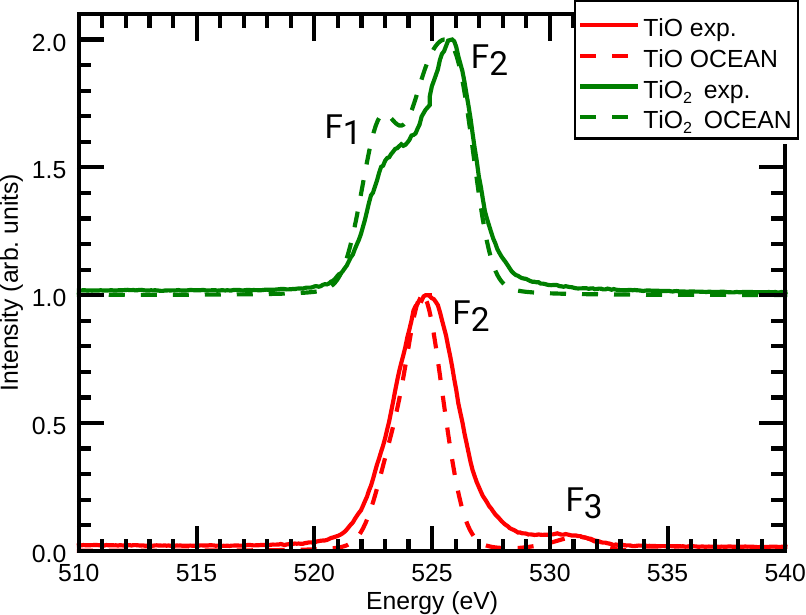}\quad
    \caption{Comparison between measured (solid lines) and calculated (dashed lines) spectra of O XES for TiO and TiO$_2$. The spectra are normalized to the respective maximum intensity and offset vertically for clarity.}
    \label{fig:o_k_xes}
\end{figure}

Fig. \ref{fig:o_k_xes} shows the XES measurement and {\sc ocean} calculation results at the oxygen K edge for TiO and TiO$_2$. Titanium oxide has a dominant feature which can be associated with the 2t$_{u}\longrightarrow$1s transition as well as the 3t$_{1u}$ transition. However, a high-energetic satellite line is observable around 530.6 eV emission photon energy, which has been seen previously in the literature \cite{valjakka_1985} and has been assigned to possible transition metal oxide transitions from the d-band. The titanium dioxide XES reveals an additional low-energy shoulder that can be associated with 2t$_{1u}\longrightarrow$1s transitions. Both are correctly represented in the {\sc ocean} calculations, with a slight overestimation of the chemical shift for TiO$_2$ which can again be associated with wrong mixing of O 2$p$ and Ti 3$d$ states.

\section{Conclusion}
We demonstrate the broad applicability of the {\sc ocean} code by providing an extensive study of titanium and two of its oxides at the K and L edges in absorption and emission spectroscopy. In this study, measurements from different beamlines with use of different spectrometers were compared. Table \ref{table:summary} summarizes the relevant comparisons between peak positions and energy differences between peaks. It presents a very good agreement for the compounds around the K-edge XAS, including the pre-edge features, the L-edge XES, and oxygen XES. The results at the L-edge XAS as well as the oxygen K-edge show slight discrepancies in the energy alignment, which might be due to wrong mixing of O 2$p$ and Ti 3$d$ states. However, the energy shifts at the L-edge provide less information than at the K edge, where the chemical shift indicates the oxidation state of the metal. More importantly are the relative peak intensities, where the calculation correctly represents the branching ratio of the L$_{II,III}$ edges.

The unique aspect of the presented comparison is the reliability of the experimental data that is used for the validation of the {\sc ocean} calculations. The use of calibrated instrumentation as well as the calibration of the energy scales provides a transferable set of data throughout different sources used, energy scales, and measurement techniques. On the other hand, the {\sc ocean} code is able to make broad predictions for all of these measurements based on the same set of input parameters, including the atomic structure and pseudopotentials. Further investigations should focus on including evolved DFT calculations and vibrational disorder to the system to reduce the remaining discrepancies.

\section*{Conflicts of interest}
There are no conflicts to declare.

\section*{Acknowledgments}
Part of this research was performed within the EMPIR project AEROMET II. This project has received funding from the EMPIR Programme co-financed by the Participating States and from the European Union's Horizon 2020 Research and Innovation Program.
Certain commercial materials are identified in this paper in order to specify the experimental procedure adequately. Such identification is not intended to imply recommendation or endorsement by NIST, nor is it intended to imply that the materials or equipment identified are necessarily the best available for the purpose.

\end{document}